\documentclass[%
superscriptaddress,
 amsmath,amssymb,
longbibliography,
prx,
twocolumn,
]{revtex4-1}
\usepackage[export]{adjustbox}
\usepackage{graphicx}
\usepackage{color}
\usepackage{xcolor}
\usepackage{wasysym}
\usepackage{dcolumn}
\usepackage{bm}
\usepackage{hyperref}
\usepackage[mathscr]{euscript}
\usepackage{mathtools}
\graphicspath{{fig/}}
\definecolor{mygreen}{rgb}{0,0.5,0}
\definecolor{mybrown}{rgb}{0.65,0.16,0.16}
\definecolor{darkpurple}{rgb}{0.494,0.184,0.556}

\def\beq {\begin{equation}}
\def\eeq {\end{equation}}
\def\beqa {\begin{eqnarray}}
\def\eeqa {\end{eqnarray}}
\def \bnum {\begin{enumerate}}
\def \enum {\end{enumerate}}
\def\bi {\begin{itemize}}
\def\ei {\end{itemize}}
\def \bdes {\begin{description}}
\def \edes {\end{description}}
\def\rel {R_{\lambda}}
\def\delx {\Delta x}
\def\delxnr {{\Delta x}/\eta}
\def\deltnr {{\Delta t}/\tau_{\eta}}
\def\dho{\partial}
\def\la {\langle}
\def\ra {\rangle}

\def\mbf {\mathbf}

\def\nabv{\mbf{\nabla}}
\def\uu{\mbf{u}}
\def\uv{\mbf{v}}
\def\ux{\mbf{x}}
\def\ur{\mbf{r}}

\def\uf{\mbf{f}}
\def\lap{{\nabla}^2}
\def\uv{\mathbf{v}}

\def\dul{\delta u}
\def\dut{\delta v}
\def\dutv{\delta \uv}
\def\duu{\delta \uu}
\def\epsm{\bar{\epsilon}}

\def\nr{r/\eta}
\def\rl{r/L}
\def\bla {\Big \langle}
\def\bra {\Big \rangle}
\def\hstar {h_\ast}
\def\hmin {h_{\textrm{min}}}
\def\snm {S_{(n,m)}(r)}
\def\snmr {S_{(n,m)}(\mbf{r})}
\def\prgl {\mathcal{P}_L}
\def\znm {\zeta_{(n,m)}(r)}
\def\xnl {\xi_{(n,0)}(r)}

\begin{document}
\title{
Scaling exponents saturate in three-dimensional isotropic turbulence
}
\author{Kartik P. Iyer}
\affiliation{
Department of Mechanical and Aerospace Engineering, New York University, 
New York, NY, $11201$, USA
}
\author{Katepalli R. Sreenivasan}
\email{krs3@nyu.edu}
\affiliation{
Department of Mechanical and Aerospace Engineering, New York University, 
New York, NY, $11201$, USA
}
\affiliation{
Department of Physics and the Courant Institute of Mathematical Sciences, New York University, New York,
NY $11201$, USA
}
\author{P. K. Yeung}
\affiliation{
Schools of Aerospace and Mechanical Engineering, Georgia Institute of Technology, Atlanta, GA $30332$, USA
}

\date{\today}
\begin{abstract}
From a database of direct numerical simulations of homogeneous and isotropic turbulence, generated in periodic boxes of various sizes, we extract the spherically symmetric part of moments of velocity increments and first verify the following (somewhat contested) results: the $4/5$-ths law holds in an intermediate range of scales and that the second order exponent over the same range of scales is {\it{anomalous}}, departing from the self-similar value of $2/3$ and approaching a constant of $0.72$ at high Reynolds numbers. We compare with some typical theories the dependence of longitudinal exponents as well as their derivatives with respect to the moment order $n$, and estimate the most probable value of the H\"older exponent. We demonstrate that the transverse scaling exponents saturate for large $n$, and trace this trend to the presence of large localized jumps in the signal. The saturation value of about $2$ at the highest Reynolds number suggests, when interpreted in the spirit of fractals, the presence of vortex sheets rather than more complex singularities. In general, the scaling concept in hydrodynamic turbulence appears to be more complex than even the multifractal description.
\end{abstract}
\pacs{Valid PACS appear here}
\maketitle
\section{Introduction} \label{intro.sec}
Velocity increments across specified separation distances are important theoretical objects in studies of three-dimensional turbulence \cite{K41a}. Their properties have been explored in a large number of papers in the past $80$ or so years and the more important results are summarized in \cite{Fri95,SA97,Ishihara09,Benzi15,ISY19}. Analogues of velocity increments have also found interesting applications in other fields such as fracture mechanics \cite{Vernede2015}, optical waves \cite{Solli07} and foreign exchange rates in financial markets \cite{FX96}. Velocity increments in turbulence, and their analogues in other fields such as those just mentioned, exhibit intense fluctuations, possess fat-tailed distributions and are typically not space-filling. The consensus of results is that the velocity increments depart from classical self-similarity that was assumed to prevail at the time of the seminal work in \cite{K41a,Obukhov41a,Obukhov41b,K41c,Heisenberg1948,Weiz1948,Onsager1949}. However, there do exist occasional claims that departures from self-similarity are artifacts of finite Reynolds numbers, and so will vanish in the limit of very large Reynolds numbers under ideal circumstances \cite{Qian1998,Arenas06,Mccomb14,Tang19,Antonia19}. Part of the reason for these latter claims is that the `consensus results' are often based on data at modest Reynolds numbers, or complicated by remnant anisotropies, or adopt Taylor's hypothesis (using time traces as one dimensional longitudinal cuts through three-dimensional fields), or employ the so-called extended self-similarity (ESS) or its variants (i.e., plotting various moments of velocity increments against the third-order), etc. These factors introduce uncertainties which, though believed to be benign, are not quantifiable precisely, and so lead to occasional divergence of conclusions. But the problem is of tremendous significance to the theory of turbulence \cite{Monin1975} to be left in this ambiguous and inconstant state.

We have recently accumulated large databases of homogeneous and isotropic turbulence in large periodic boxes (as large as $16,384^3$ grid points) with good resolution in both space and time \cite{YS2005,DY10,YSP18}, which can be used to assess the status of fundamental issues such as self-similarity, intermittency, and universality. There is an acceptable likelihood of reaching satisfactory conclusions because: The highest Taylor microscale Reynolds number of these data is $1300$, which appears to be high enough to expect decent scaling (for Eulerian quantities) without the need for ESS or its variants; there is obviously no need for Taylor's hypothesis because the data are spatial; we successfully remove by spherical averaging the residual anisotropies inherited from forcing and the cubic shape of the simulations box; and the amount of statistically stationary data available is adequate for high-order moments of velocity increments to converge reliably. Indeed, the analysis of the data shows that the departures from the estimates based on self-similarity assumptions on velocity increments are real and do not vanish with increasing Reynolds number. They also show that, even in isotropic turbulence, there is a persistent difference between the longitudinal and transverse velocity increments, as was pointed out already in \cite{Chen,Dhruva}; the particular new result, which we believe is of far-reaching theoretical consequence, is that the scaling exponents of the transverse increments saturate for high-order increment moments reminiscent of Burgers turbulence and passive scalars \cite{Gotoh94,CLMV2000,GF01,Staicu03,KI18}. We will briefly examine the reason why. The longitudinal moments might also saturate, but, if that does happen, it would do so for moments of far higher order---the ones that cannot be computed reliably. 

Section \ref{method.sec} presents an account of numerical methods and flow parameters, as well definitions for later use. The account of numerical methods is necessarily brief because a more detailed description can be found in Refs.~\cite{YDS2012,Yeung15}. Section \ref{res.sec} presents the bulk of the results and is followed by discussions and conclusions in Sec.~\ref{conc.sec}. 

\section{Numerical method, flow parameters and definitions}
\label{method.sec}

We solve the incompressible Navier-Stokes equations in three-dimensions,  
\beq
\label{nse.eq}
{\dho\uu}/{\dho t} + \uu\cdot \nabv \uu = -\nabv p + \nu \lap \uu + \uf \;,
\eeq
by using direct numerical simulations (DNS) on a triply periodic $N^3$ box with edge-length $L_0$, where $\uu(\ux,t)$ is the solenoidal velocity field ($\nabv\cdot \uu = 0$), $\nu$ is the kinematic viscosity and $p$ is the kinematic pressure, and $\uf$ the large-scale forcing in the range $r_f \in (0.2, 0.5) L_0$ (technically, in the corresponding wavenumber range) \cite{DY10}. We use the standard pseudospectral scheme with exponential convergence, and calculate the nonlinear terms in physical space. The time-stepping is done with an explicit second-order Runge-Kutta integration to evolve the flow to a statistically stationary state to which all the present results correspond. The results have been averaged over a stationary period of at least $10$ large-eddy time scales $L/u^\prime$, where $L \approx 0.2 L_0$ is the so-called integral scale and $u^\prime$ is the root-mean-square velocity fluctuation. Both $L$ and $u^\prime$ are independent of viscosity or, equivalently, the Reynolds number. We present results in terms of the microscale Reynolds number $\rel \equiv u^\prime \lambda/\nu$, where the Taylor microscale $\lambda$ is given by $u'/\sqrt{\la(\dho u/\dho x)^2\ra}$. The spatial resolution $\Delta x/\eta$, where the grid spacing $\Delta x = L_0/N$ and the Kolmogorov scale $\eta \equiv (\nu^3/\epsm)^{1/4}$, given in terms of the mean energy dissipation rate $\epsm$, is listed in Table \ref{dns.tab}, along with the other relevant flow parameters. In Appendix \ref{numres.app} we have verified that the results provided in this paper are consistent with those from shorter simulations \cite{YSP18} that used finer spatial and temporal resolution.  

\setlength{\tabcolsep}{16pt}
\begin{table}
\centering
\caption{Simulation parameters: $N^3$ is the number of grid points, $\rel$ is the microscale Reynolds number, $(L/\eta)^3$ is a measure of the number of degrees of freedom in the three-dimensional field, and $\delx/\eta$ is the ratio of the grid spacing to the Kolmogorov scale. 
}
\label{dns.tab}
\begin{tabular}{cccc}
\\
\hline
$N^3$ & $\rel$ & $L/\eta$ & $\delx/\eta$\\
\hline
$256^3$ & $140$ & $108$ & $2.1$ \\
$512^3$ & $240$ & $226$ & $2.1$ \\
$2048^3$ & $400$ & $446$ & $1.1$ \\
$4096^3$ & $650$ & $898$ & $1.1$ \\
$8192^3$ & $650$ & $909$ & $0.6$ \\
$8192^3$ & $1300$ & $2514$ & $1.5$ \\
$16384^3$ & $1300$ & $2522$ & $0.8$ \\
\hline
\end{tabular}
\end{table}

We define a few parameters for later use in the paper. Consider the two-point velocity increment at location $\ux$ across a separation vector $\ur$ with magnitude $r \equiv |\ur| > 0$, $\duu(\ux,\ur) = \uu(\ux+\ur)-\uu(\ux)$. Define the longitudinal increment $\dul(\ux,\ur) = \duu(\ux,\ur)\cdot\hat{\ur}$, where $\hat{\ur} = \ur/r$ is the unit vector along $\ur$, and the transverse increment vector $\dutv(\ux,\ur) = \duu(\ux,\ur)-\dul(\ux,\ur)\hat{\ur}$. The magnitudes are written, for the transverse case as an example, as $\dut(\ux,\ur) = |\dutv(\ux,\ur)|$. The velocity increment moment, also known as the structure function, $\snm$ at order $n+m$, is defined as
\beq
\label{snm.eq}
\snm \equiv \la (\dul)^n (\dut)^m \ra \;,
\eeq
where $\la \cdot \ra$ denotes space, time and angle (or spherical) averages \cite{TKE03,KI17}. The angle averaging is performed to obtain the isotropic sector of $\snmr$ from its SO(3) expansion \cite{Kurien00,BP05}. This step is necessary to eliminate any residual anisotropy effects that may be present due to the specific method of forcing at low wave numbers and the geometry of the box. At sufficiently large $\rel$, if there exists an inertial range of scales that are smaller than the integral scale $L$ (where energy injection occurs) and larger than the viscous scales $\sim \eta$ (where dissipation manifests), the structure functions in that range are expected to scale as $\snm \sim (r/L)^{\zeta_{(n,m)}}$, where $\zeta_{(n,m)}$ are the scaling exponents. For later use we note that the self-similar scaling attributed to Kolmogorov \cite{K41a} gives $\zeta_{n+m}^{k41} = (n+m)/3$.

\subsection{Third order structure function}
\label{thirdord.sec}

\begin{figure}
\begin{minipage}{0.5\textwidth}
\includegraphics [width=1.0\textwidth,center]{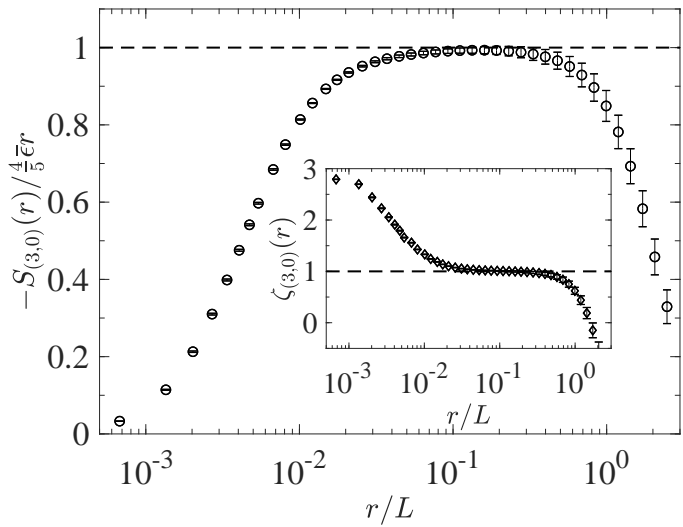}
\end{minipage}
\protect\caption{
Compensated third-order isotropic longitudinal structure function versus scale $r$ on log-linear scales. The maximum value of $[-S_{(3,0)}(r)/(\frac{4}{5})\epsm r] = 0.99\pm 0.01$. Inset shows the logarithmic local slope $\zeta_{(3,0)}(r) = d[log|S_{(3,0)}(r)|]/d[\log r]$. Dashed line at unity, in both the main figure and the inset, is the exact result of Kolmogorov (see Eq.~\ref{k41.eq}). 
}
\label{dlll.fig}
\end{figure}

We shall first consider structure function of order $3$, for which an exact result has been derived from Eq.~\ref{nse.eq} in the inertial range (if one exists). This so-called Kolmogorov's $4/5$-ths law \cite{K41c} is given by
\beq
\label{k41.eq}
S_{(3,0)}(r) = -\frac{4}{5} \epsm r.
\eeq
Figure \ref{dlll.fig} shows that, at $\rel=1300$, Eq.~\ref{k41.eq} is satisfied within error bars in the range $\rl \in (0.05,0.4)$, to the sides of which dissipative or large scale effects manifest to produce deviations from Eq.~\ref{k41.eq}. The logarithmic local slope of $S_{(3,0)}$ given in the inset of Fig.~\ref{k41.eq} shows excellent agreement with the power-law exponent of unity in Eq.~\ref{k41.eq}. The angle-averaged result shown in Fig.~\ref{dlll.fig} corresponds to the isotropic sector of the SO(3) decomposition of $S_{(3,0)}(\ur)$ and extends the inertial range by a factor of two over the Cartesian-averaged result for the same data (Fig.~$2$ of Ref.~\cite{KIKRS17}). This dispels the explicit claim of Ref.~\cite{Antonia19} that the evidence for the 4/5-ths law does not exist.

\section{Results}
\label{res.sec}

\subsection{Second order structure function and the intermittency exponent}
\label{secord.sec}

Before examining structure functions of order $2$, it is instructive to test the incompressible relation in isotropic turbulence at scale $r$, 
\beq
\label{iso2.eq}
S_{(0,2)}(r) = 2S_{(2,0)}(r) + r\frac{d}{dr} S_{(2,0)}(r).
\eeq
\begin{figure}
\begin{minipage}{0.5\textwidth}
\includegraphics [width=1.0\textwidth,center]{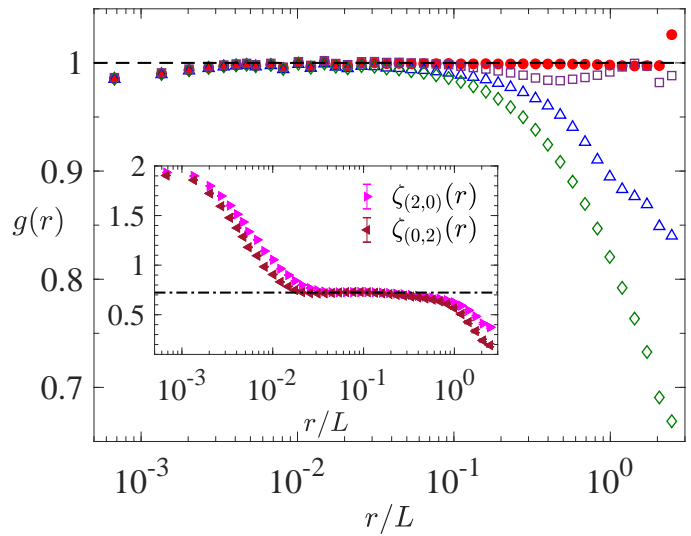}
\end{minipage}
\protect\caption{
Ratio of the right hand side to the left hand side of Eq.~\ref{iso2.eq}, $g(r)$ versus scale $r$ at $\rel=1300$, computed on a $8192^3$ periodic box. Open symbols correspond to scale separation along the Cartesian directions $\hat{\ur} = (1,0,0)$ (triangle), $\hat{\ur} = (0,1,0)$ (diamond) and $\hat{\ur} = (0,0,1)$ (square); filled circle is the isotropic sector from the SO(3) decomposition of the structure functions. If isotropy holds at scale $r$, we expect $g(r)=1$, which is marked by the dashed line at unity; the data follow this expectation approximately for $r/L > 5 \times 10^{-3}$, corresponding roughly to $r/\eta > 10$. Inset shows that the logarithmic local slopes of $S_{(2,0)}$ and $S_{(0,2)}$ for the isotropic sector are equal to each other in the range $\rl \in (0.02,0.2)$. The numerical value of $0.72$ marked by the dot-dashed line is discussed in the text and the next figure.
}
\label{iso2.fig}
\end{figure}

Figure \ref{iso2.fig} displays $g(r)$, the ratio of the right hand side to the left hand side of Eq.~\ref{iso2.eq}, calculated along the three Cartesian directions and using the isotropic sector (filled circle) of the second order structure function. The remnant anisotropy from the cubic grid geometry and large scale forcing render velocity increments along different directions anisotropic to different degrees, resulting in $g(r)$ different from unity even at moderately small scales. But the angle averaging, which retains only the isotropic sector, guarantees that $g(r) = 1$ at almost all scales, as the figure clearly shows. If the longitudinal and transverse second-order structure functions display power-law behaviors in the inertial range, Eq.~\ref{iso2.eq} implies that the exponents $\zeta_{(2,0)}$ and $\zeta_{(0,2)}$ must be equal. The inset of Fig.~\ref{iso2.fig} verifies this expectation.
\begin{figure}
\begin{minipage}{0.5\textwidth}
\includegraphics [width=0.89\textwidth,center]{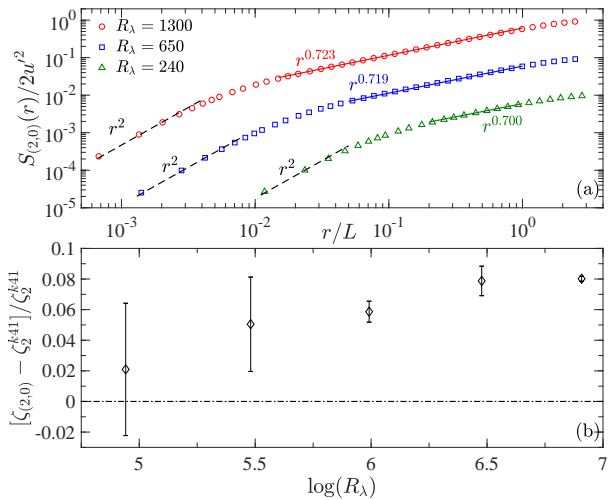}
\end{minipage}
\protect\caption{(a) Scaling of the second-order structure function normalized by $2{u^\prime}^2$ such that $S_{(2,0)}(L)/2{u^\prime}^2 \approx 1$, where $u^\prime$ is the root-mean-square velocity fluctuation. For clarity, the curves for $\rel = 650$ and $240$ are shifted below by a factor of $10$ and $100$, respectively. Least-square fits in the inertial range are shown. Dashed lines with slope $2$ are shown at smallest scales to verify viscous scaling. (b) Relative deviation of the second-order inertial range exponent from the Kolmogorov value of $2/3$ versus the logarithm of the microscale Reynolds number $\rel$. Dashed line at zero corresponds to Kolmogorov scaling of $2/3$. Vertical bars indicate the standard error due to temporal variations in the least-square fits. The exponents approach a constant $\zeta_{2,0} = 0.72~\pm~0.004$ corresponding to a constant, $\rel$-independent correction at higher $\rel$. 
}
\label{sf2.fig}
\end{figure}

Since the second-order exponents $\zeta_{2,0}=\zeta_{0,2}$, as shown in the inset of Fig.~\ref{iso2.fig}, it suffices to examine more closely the longitudinal structure function (say), as is done in Fig.~\ref{sf2.fig}. The second-order structure function displays proper power-laws in the expected scale range (see inset of Fig.~\ref{iso2.fig}). The deviation of the exponent from the self-similar Kolmogorov exponent, obtained by least-squares, is plotted in Fig.~\ref{sf2.fig}(b) as a function of log $R_\lambda$. The correction increases for lower $R_\lambda$ but saturates at higher $\rel$ at a constant value of about 0.72, about $8\%$ higher than $2/3$. 

\subsection{Isotropy and fourth order quantities}
\label{fourord.sec}

Exact dynamical equations derived for isotropic structure functions of even orders \cite{Yakhot01,Hill2001} contain mixed-order structure functions and structure functions of pressure and velocity increments. The equation for the longitudinal structure functions of order $2n$ is given by
\beqa
\label{long.eq}
\frac{\dho S_{(2n,0)}}{\dho r}+\frac{2}{r}S_{(2n,0)} &=& 
\frac{(2n-1)}{r} S_{(2n-2,2)} \\
\nonumber 
&-&(2n-1) \la \prgl (\dul)^{2n-2} \ra \\
\nonumber 
&+& \epsm[1-\cos(r/r_f)] a_n S_{(2n-3,0)}
\eeqa
where $a_n = 2(2n-1)(2n-2)/3$ and $\prgl \equiv (\nabv p(\ux+\ur)-\nabv p(\ux)).\hat{r}$ is the longitudinal pressure gradient structure function. Similar equations relating the transverse and mixed structure functions are also known \cite{Yakhot01,Hill2001,KS01}. The pressure contributions were initially thought to be small in the inertial range \cite{Yakhot01,KS01}, which led to the result that the scaling exponents for a given order are equal, i.e., $\zeta_{(2n,0)} = \zeta_{(2n-2,2)} = \zeta_{(0,2n)}$ for $2n \ge 4$.
\begin{figure}
\begin{minipage}{0.5\textwidth}
\includegraphics [width=1.0\textwidth,center]{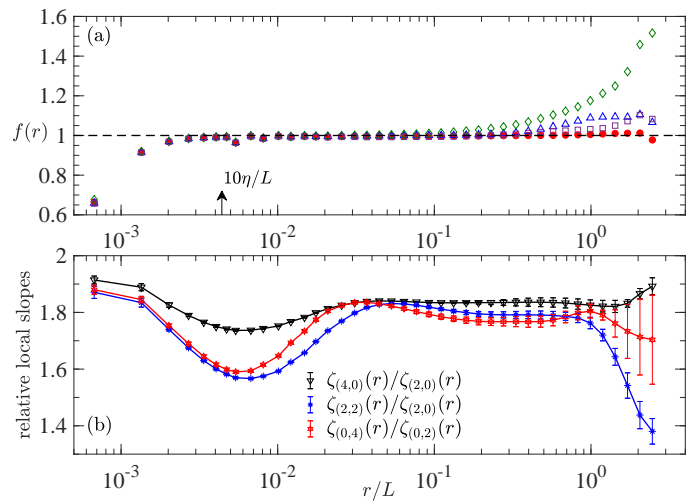}
\end{minipage}
\protect\caption{
(a) Examination of isotropy of moment order four. Ratio of the right hand side to the left hand side of Eq.~\ref{sf4.eq}, $f(r)$ versus $r$ for $\rel=1300$. Open symbols correspond to scale separation along Cartesian directions $\hat{\ur} = (1,0,0)$ (triangle), $\hat{\ur} = (0,1,0)$ (diamond) and $\hat{\ur} = (0,0,1)$ (square); filled circles are for the isotropic sector from the SO(3) decomposition of structure functions. If isotropy holds at scale $r/\eta \gg 1$, we should have $f(r)=1$ (marked by the dashed line), as the filled circles show to be true with no ambiguity beyond $r/\eta = 10$, marked on the abscissa for reference. (b) Ratio of the logarithmic derivatives of the fourth order structure functions relative to the second order structure functions versus $r$ for the same data. The self-similar, intermittency-free value for this ratio is $2$. Thus, even when isotropy is guaranteed, departures from the self-similar Kolmogorov-scaling prevail.
}
\label{sf4.fig}
\end{figure}

In order to examine isotropy persuasively we substitute $2n=4$ in Eq.~\ref{long.eq} to yield the following exact isotropic relation valid for $r/\eta \gg 1 $:
\beq
\label{sf4.eq}
\frac{\dho S_{(4,0)}}{\dho r}+\frac{2}{r}S_{(4,0)} = 
\frac{3}{r} S_{(2,2)}
- 3\la \prgl (\dul)^{2} \ra.
\eeq
Here the large scale contributions drop out because $S_{(1,0)}=0$. Figure \ref{sf4.fig}(a) shows the ratio $f(r)$ of the right hand side to the left hand side of Eq.~\ref{sf4.eq}, calculated along the three Cartesian directions and for the isotropic sector (filled circle) of the fourth order structure functions. The isotropic sector does indeed satisfy Eq.~\ref{sf4.eq} beyond $r/\eta = 10$ exceedingly well. This result ensures isotropy at order four for $S_{4,0}$ and $S_{2,2}$ for all $r/\eta > 10$ . 

We now examine the scaling exponents $\znm = d[\log\snm]/d[\log r]$ for order four and assess the contributions of pressure. Figure \ref{sf4.fig}(b) compares the ratios $\zeta_{(4,0)}/\zeta_{(2,0)}$ and $\zeta_{(2,2)}/\zeta_{(2,0)}$, which are essentially constant in the inertial range (approximately in the region $0.1 < r/L < 1$), suggesting that $S_{(4,0)}$ and $S_{(2,2)}$ display power-laws over this region. Also shown for comparison is the ratio $\zeta_{(0,4)}/\zeta_{(0,2)}$, noting that $\zeta_{(2,0)} = \zeta_{(0,2)}$ due to incompressibility (see Fig.~\ref{iso2.fig}(b)). The exponents $\zeta_{(4,0)}$ and $\zeta_{(2,2)}$ show non-trivial differences in this range which indicate that the pressure contribution to Eq.~\ref{sf4.eq} is not negligible --- at least for this Reynolds number. 
\subsection{Further comments on pressure contributions}\label{pr.sec}

\begin{figure}
\begin{minipage}{0.5\textwidth}
\includegraphics [width=1.0\textwidth,center]{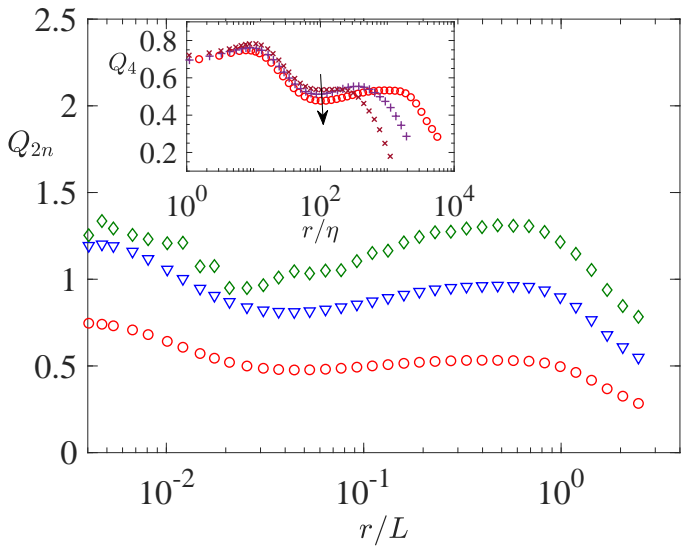}
\end{minipage}
\protect\caption{
The non-dimensional quantity $Q_{2n}$, which is the ratio of the pressure contribution in Eq.~\ref{long.eq} to the longitudinal structure function, plotted against $r/L$ for orders $2n = 4$ (circle), $6$ (triangle) and
$8$ (diamond) at $\rel = 1300$. The inset shows $Q_4$ versus $\nr$ at three different Reynolds numbers: $\rel = 400$ (cross), $\rel = 650$ (plus) and $1300$ (circle). The pressure contributions that cause the differences between $S_{4,0}$ and $S_{2,2}$ seem to decrease with increasing $\rel$ as indicated by the arrow in the inset, but this appears to happen, if at all, very slowly (consistent with \cite{iyeretal2019}).
}
\label{sflprdelu.fig}
\end{figure}
  
In order to ascertain the role of pressure in Eq.\ \ref{long.eq} we plot in Fig.~\ref{sflprdelu.fig} the ratio of the pressure term to that of the longitudinal structure function for orders $2n = 4,6$ and $8$ at $\rel = 1300$; here $Q_{2n} = -(2n-1)r \la \prgl (\dul)^{2n-2} \ra/S_{(2n,0)}$. With increasing order, the ratio $Q_{2n}$ increases in the intermediate scale range, with the gap between successive orders decreasing, perhaps suggesting that they level off to some non-zero value for some high orders not accessible to measurement today. A similar examination of ratios $r \la \prgl (\dul)^{2n-2} \ra /S_{(2n-2,2)}$ shows the same qualitative behavior. The conclusion is that the pressure effects between longitudinal and mixed structure functions increase upscale (see the green diamonds in Fig.~\ref{sflprdelu.fig}) and with increasing order (though perhaps less rapidly), leading to the persistent differences observed in Fig.~\ref{sf4.fig}(b) between $\zeta_{(4,0)}$ and $\zeta_{(2,2)}$. The inset compares the non-dimensional ratio $Q_4$ at three different Reynolds numbers against scale $r$ normalized by $\eta$, for purposes of examining the Reynolds number dependence of pressure contributions. The ratio $Q_4$ decreases slowly with increasing Reynolds number causing the exponents to come closer \cite{SW02,KIKRS17}. It follows that, for a given finite Reynolds number, the pressure contributions differentiate between the exponents $\zeta_{(2n,0)}$ and $\zeta_{(2n-2,2)}$. On the other hand, it is well known that the pressure effect on the transverse structure functions is markedly smaller \cite{KS01,Gotoh2003} and, in fact, decreases upscale, causing the mixed and the transverse exponents $\zeta_{(2,2)}$ and $\zeta_{(0,4)}$ to approach each other at smaller scales, as seen in Fig.~\ref{sf4.fig}(b).
\begin{figure}
\begin{minipage}{0.5\textwidth}
\includegraphics [width=1.0\textwidth,center]{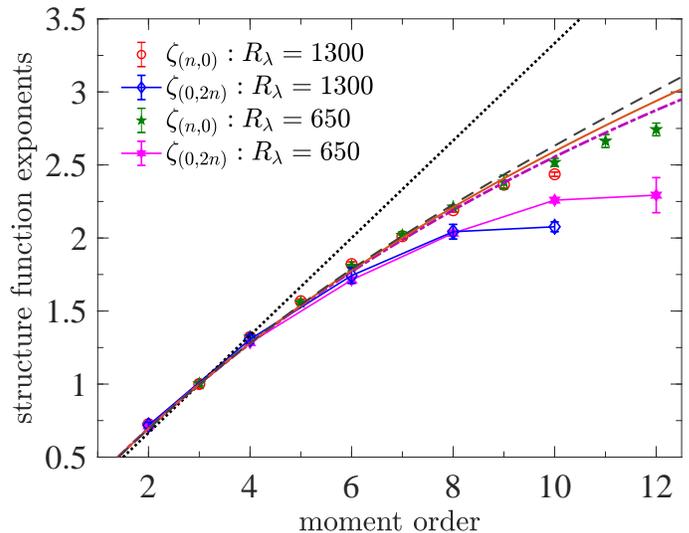}
\end{minipage}
\protect\caption{Scaling exponents of longitudinal structure functions $\zeta_{(n,0)}$ and even-order transverse structure functions $\zeta_{(0,2n)}$ as functions of moment order at $\rel = 1300$ (open symbols) and $\rel = 650$ (filled symbols). Dash-dot curve is the model prediction by Yakhot \cite{Yakhot01}, solid curve is the She-Leveque model \cite{SL94} while the dashed line is the $p$-model by Meneveau and Sreenivasan \cite{CMKRS87}. Dotted line is the intermittency free, self-similar result of Kolmogorov \cite{K41a}. Error bars indicate $95\%$ confidence intervals. The transverse exponents saturate at $\zeta^T_\infty \approx 2.2$ at $\rel = 650$, while they seem to saturate at about $2$ at $\rel = 1300$.
}
\label{velexp.fig}
\end{figure}

\subsection{Saturation of higher order exponents}
\label{sat.sec}
We summarize in Fig.~\ref{velexp.fig} the exponents for integral orders. Focusing first on the longitudinal exponents $\zeta_{(n,0)}$, the data extend to $n = 12$ for $\rel = 650$ but had to be truncated at $n = 10$ for $\rel=1300$ for reasons of statistical convergence; a brief assessment of the convergence of moments is presented in the next section. The longitudinal data for the two Reynolds numbers shown agree with each other for $n<8$, beyond which they begin to differ modestly; we are not certain that the differences represent genuine Reynolds-number effects and will not focus on those modest differences here. The longitudinal exponents appear to closely follow the model by Yakhot \cite{Yakhot01,JS2007} (almost up to order $10$) while they increasingly deviate from the She-Leveque model \cite{SL94} and the $p$-model \cite{CMKRS87} at higher orders. (The data differ also from an interesting model by \cite{Ruelle12} but we do not show this comparison here because the model does not preserve the concavity property of the exponents.)

Also shown in the figure are the even-order exponents for transverse structure functions. These exponents agree with the longitudinal data for moment orders approximately up to $n = 4$. Beyond that, for higher orders, focusing first on $\rel = 650$, the transverse exponents show a tendency to saturate to a value $\zeta^T_\infty$ of about $2.2$. For the higher $\rel$, the tendency to saturate begins at lower moment order (perhaps $6$), and the saturation value appears to be about $2$ giving $\zeta^T_\infty \approx 2$. 
\begin{figure}
\begin{minipage}{0.5\textwidth}
\includegraphics [width=1.0\textwidth,center]{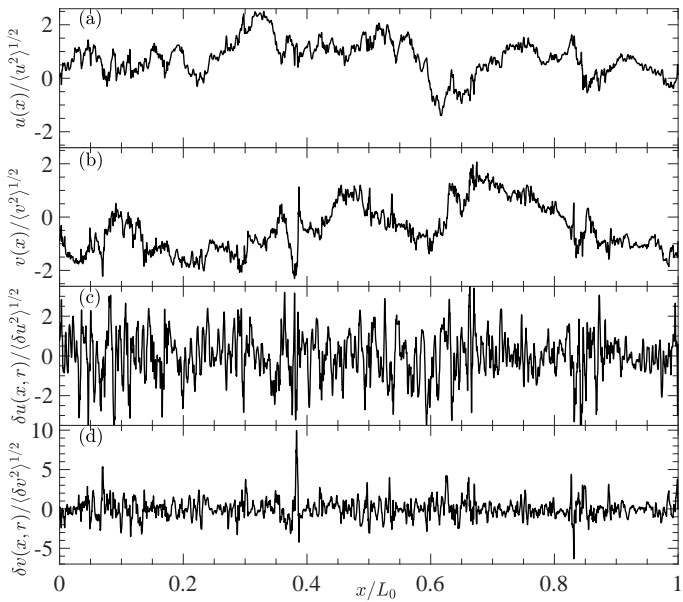}
\end{minipage}
\protect\caption{
Velocity fluctuation components 
$u(x,y_0,z_0) \equiv \uu \cdot \hat{\mathbf{r}}_1$, $\hat{\mathbf{r}}_1=(1,0,0)$ and $v(x,y_0,z_0) \equiv \uu \cdot \hat{\mathbf{r}}_2$, $\hat{\mathbf{r}}_2=(0,1,0)$ are shown in panels (a) and (b) as functions of the spatial co-ordinate $x$ for fixed $(y_0,z_0)$ in a cube with side $L_0$ computed at a resolution $8192^3$, $R_\lambda = 1300$. Panels (c), (d) show the longitudinal and transverse velocity difference traces corresponding to panels (a), (b), with $\ur = r \hat{\mathbf{r}}_1$ and $\ur = r \hat{\mathbf{r}}_2$, respectively, for $r/L_0 = 0.004$ ($r/\eta = 47$) as functions of the spatial co-ordinate $x$. Traces in all panels are normalized by their respective standard deviations. Occasional spikes in transverse increments (see near $x/L_0 \approx 0.39$) do not appear in the longitudinal increments. Also compare the traces themselves in (a) and (b) for the same spatial position.
}
\label{traces.fig}
\end{figure}

The saturation of transverse exponents suggests that there must be some huge excursions in transverse increments, unlike in the longitudinal counterparts, that imprint their characteristics on high-order structure functions. This is clearly seen in the traces provided in Fig.~\ref{traces.fig}. In turbulence, the nonlinear effects that steepen gradients are balanced by the effect of pressure that mitigates it. What we have seen is that the pressure effect on transverse velocity increments is weak, giving rise to steeper structures in those signals. Thus, transverse exponents saturate and exhibit a greater degree of intermittency than longitudinal increments \cite{RHK91,Gotoh2003}. Since the pressure effect seems to fade very slowly at a fixed order with increasing $\rel$ (see inset in Fig.~\ref{sflprdelu.fig}), it is possible that the longitudinal exponents may also saturate in the limit $\rel \to \infty$. Indeed, Yakhot's model predicts that even longitudinal exponents eventually saturate at $7.66$ as $n \to \infty$, but its verification is beyond the capabilities of the present data (or the foreseeable ones, since a simple extrapolation suggests that it would require Reynolds numbers beyond those occurring on Earth). Indeed the estimation of scaling exponents at large orders, requiring immense amounts of data, may be affected by the spurious effects of the sort discussed in \cite{lashermes04}. What appears certain from the $\rel$-trend in Fig.~\ref{velexp.fig} is that the higher order transverse exponents saturate.

\subsection{Statistical convergence of moments and tails of probability density functions}
\label{conv.sec}
\begin{figure}
\begin{minipage}{0.5\textwidth}
\includegraphics [width=1.0\textwidth,center]{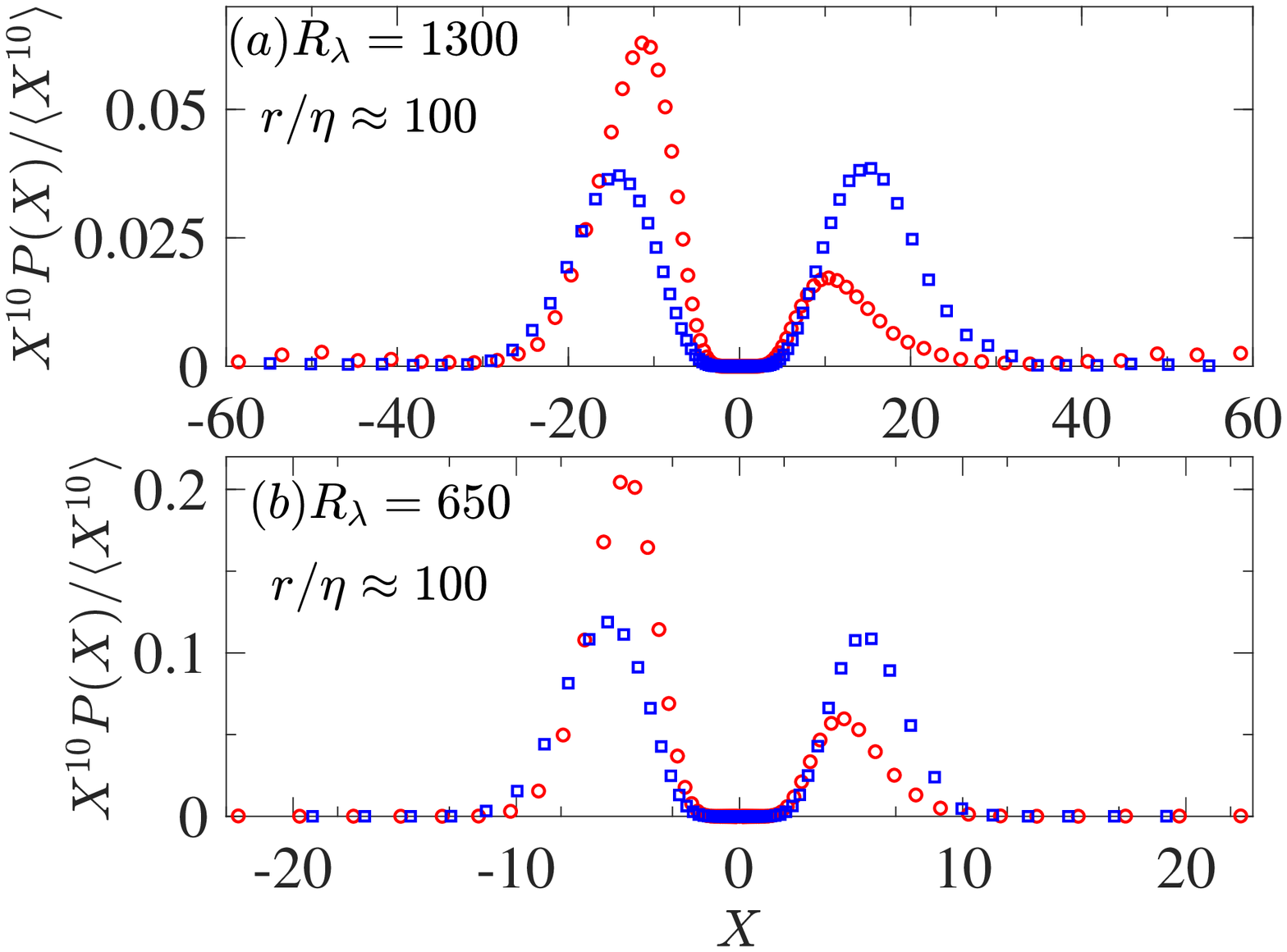}
\end{minipage}
\protect\caption{
Tenth order integrand of normalized longitudinal velocity increment (circle) $X \equiv \dul/\la \dul^2 \ra^{1/2}$ and normalized transverse increment (square) $X \equiv \delta v(\hat{\ur}) /\la {\delta v(\hat{\ur})}^2 \ra^{1/2}$ [$\uv \cdot \hat{\ur} = 0$ with $\hat{\ur} = (1,0,0), (0,1,0)$ and $(0,0,1)$] at $\nr \approx 100$ (approximately the ultraviolet end of the inertial range). (a) The calculations for $\rel = 1300$ are on a $8192^3$ grid and (b) those for $\rel = 650$ on a $4096^3$ grid. The integrands are normalized by respective moments such that the areas under each curve sum up to unity.
}
\label{conv.fig}
\end{figure}

The statistical convergence of the higher order moments of the longitudinal and transverse increments is confirmed by the rapid decay towards both tails of the moment integrands as shown in Fig.~\ref{conv.fig}. The integrands of moment orders $10$ for $\rel=1300$ and $\rel=650$ are shown in Fig.~\ref{conv.fig}, each at the lower end of the inertial range. The integrands peak well before the tail contributions decay, ensuring statistical convergence of the moments.

\begin{figure}
\begin{minipage}{0.5\textwidth}
\includegraphics [width=1.0\textwidth,center]{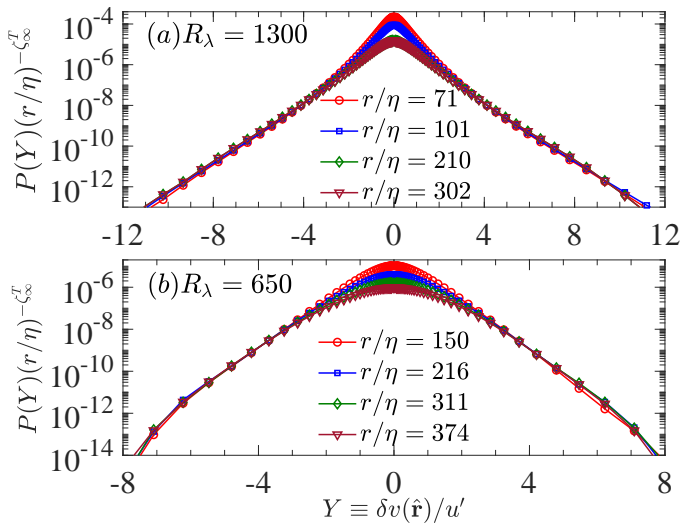}
\end{minipage}
\protect\caption{
Probability density function $P$ of the transverse velocity increment $\delta v(\hat{\ur})$ [$\uv \cdot \hat{\ur} = 0$ with $\hat{\ur} = (1,0,0), (0,1,0)$ and $(0,0,1)$], in the inertial range, normalized by the root-mean-square velocity fluctuation, compensated by $r^{-\zeta^T_\infty}$, where $\zeta^T_\infty$ is the transverse saturation exponent from Fig.~\ref{velexp.fig}. (a) $\rel = 1300$: $\zeta^T_\infty = 2.05$, (b) $\rel = 650$: $\zeta^T_\infty = 2.2$. The compensated tails of $P$ collapse supporting the saturation of transverse exponents. 
}
\label{sat.fig}
\end{figure}
Saturation of transverse exponents at higher orders implies that present in the transverse velocity increments are jumps $\dut \gtrsim u^\prime$, which implies that the tails of the probability density function 
$P(\dut) \propto r^{\zeta_\infty^T}$. Figure \ref{sat.fig} verifies that this is indeed the case with $P[\delta v(\hat{\mathbf{r}})] r^{-\zeta_\infty^T}$ collapsing for $\dut > 3u^\prime$, across the inertial separations. In contrast, the compensated probability density functions of the longitudinal increments do not collapse in this fashion even for $\rel=1300$. 

\subsection{Exponent derivatives}
\label{ord.sec}
We now consider the derivatives of absolute values of structure functions of various orders. Define for longitudinal quantities the local slope of order-$n$ as
\beq
\label{ls.eq}
\xnl \equiv \frac{d}{d\log r}[\log\la |\dul|^n \ra] \;,
\eeq
whose constancy in the inertial range yields the longitudinal scaling exponent at order $n$. Clearly, $\xi_{(2n,0)} = \zeta_{(2n,0)}$, but the two may differ for odd orders. Differentiating Eq.~\ref{ls.eq} with respect to $n$ we get the exponent derivative
\beq
\label{dldz.eq}
\frac{d\xi_{(n,0)}(r)}{dn} = \frac{d}{d \log r} \Bigg[ \frac{\bla |\dul|^n \log |\dul| \bra} {\la |\dul^n| \ra}\Bigg ] \;.
\eeq
In particular, the exponent-derivative for $n=0$ is 
\beq
\label{derzero.eq}
\frac{d\xi_{(n,0)}(r)}{dn} \Big |_{n=0} = 
\frac{d}{d \log r} [ {\la \log |\dul| \ra} ] \;.
\eeq
\begin{figure}
\begin{minipage}{0.5\textwidth}
\includegraphics [width=1.0\textwidth,center]{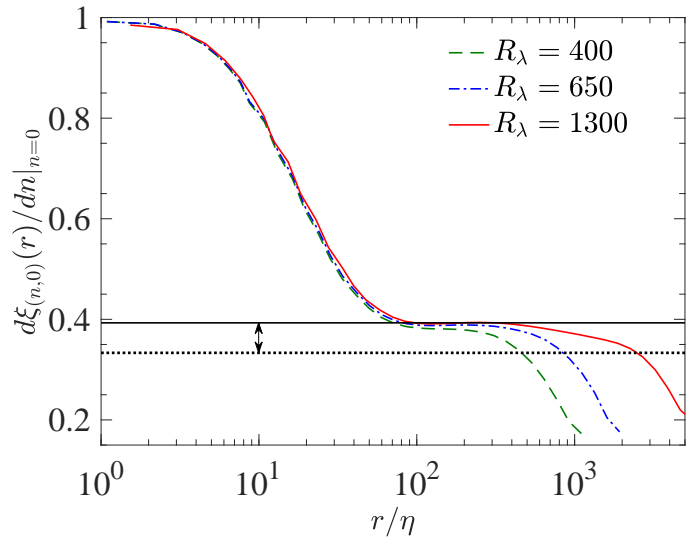}
\end{minipage}
\protect\caption{Exponent derivative of the absolute longitudinal structure function exponent $\xi_{(n,0)}$ at $n=0$ versus scale $r$ at three different Reynolds numbers. The $r \to 0$ asymptote is $1$ as expected. The inertial range plateau marked by the solid line yields $d\xi_{(n,0)}/dn|_{n=0} \approx 0.39$ from a least-squares fit for $\rel = 1300$. Horizontal dotted line at $1/3$ is the corresponding Kolmogorov value. Arrow shows the degree of anomaly at order $0$ which is close to $18\%$.   
}
\label{dzdprfun.fig}
\end{figure}

Figure \ref{dzdprfun.fig} plots the zeroth-order derivative as a function of scale at three different Reynolds numbers. In the viscous limit $r/\eta \to 1$ all curves approach unity which one expects from Taylor series expansion. In the inertial range the curves approach a scale independent plateau which is the order derivative corresponding to $n=0$. Intermittency exists even at order zero (as was noted already in Ref.~\cite{chen05}) and is seen to saturate at $0.39\pm 0.001$ for $\rel \geq 600$. In the multifractal model \cite{Fri85} this number corresponds to the most probable H\"{o}lder exponent, $\hstar(n=0)$ at which the fractal set $D(h)$ attains its maximum $D(\hstar(0))=3$.
\begin{figure}
\begin{minipage}{0.5\textwidth}
\includegraphics [width=1.0\textwidth,center]{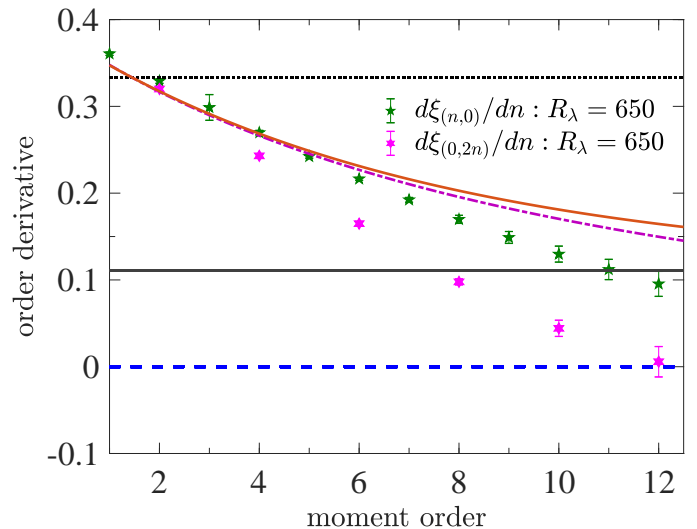}
\end{minipage}
\protect\caption{Order derivative of the absolute longitudinal $\xi_{(n,0)}$ and even-order transverse $\xi_{(0,2n)}$ structure function exponents versus moment order at $\rel = 650$. Dotted line at $1/3$ is the intermittency-free result of Kolmogorov \cite{K41a}. The dash-dot curve is that of Yakhot \cite{Yakhot01} while the solid curve corresponds to the She-Leveque model \cite{SL94}. The horizontal solid line at $1/9$ marks the asymptotic order derivative of the She-Leveque model, while the dashed line at zero corresponds to the saturation of exponents, which is also the asymptotic order-derivative for the Yakhot model.
}
\label{dzdp.fig}
\end{figure}

The exponent derivatives for $n \ge 0$ for both the absolute longitudinal and even-order transverse increments at $\rel=650$ (Eq.~\ref{dldz.eq}) are summarized in Fig.~\ref{dzdp.fig}. The derivatives decrease monotonically with the order since the exponents are concave in $n$ due to the H\"{o}lder inequality, with the longitudinal and transverse derivatives differing from order $4$ or so onward, with the latter 
dropping more steeply than the former. In the multifractal terminology, this means that the dominant H{\"o}lder exponent $\hmin(n)$ of the transverse increment is smaller than that of the longitudinal increment from order $4$ onwards rendering the transverse increments to be more intermittent than the longitudinal counterparts. The transverse order derivative reaches zero at order $12$ (consistent with the earlier finding about the saturation) whereas that of the longitudinal exponent continue to possess a positive slope which decreases with increasing order, roughly reaching the asymptotic $\hmin(n \to \infty)$ of the She-Leveque model \cite{SL94}. These conclusions are entirely consistent with Figs.~\ref{sf4.fig}(b) and \ref{velexp.fig}. Since negative scaling exponents for incompressible flows are ruled out in the multifractal model \cite{Fri95}, we note that the most dominant H{\"o}lder exponent $\hmin^T(n \to \infty)=0$ corresponding to the saturation of transverse exponents, suggesting the physical presence of shock-like structures in the transverse increment field, as seen in Fig.~\ref{traces.fig}(d). 

\section{Discussion and conclusions} \label{conc.sec}
At the time of the seminal works of Kolmogorov \cite{K41b,K41c,Obukhov41a,Obukhov41b,Heisenberg1948,Weiz1948,Onsager1949}, it was thought that fluid turbulence comprises scales that progressively degenerate in structural integrity, essentially becoming more isotropic as the energy cascade proceeds. On the basis of this scale invariance the second order velocity structure function was deduced to scale as $r^{2/3}$ --- or, equivalently, the energy spectrum as $k^{-5/3}$, where $k$ is the wavenumber. Its proper experimental verification \cite{Grant1962} had to wait many years, which also sowed the seeds for the incompleteness of the self-similar theory.  Subsequently a number of works starting from \cite{K62} have revised this earlier picture and it is now well known that turbulent structures of moderate scale tend be less space filling and intermittent, far from being spherical, causing departures from self-similarity \cite{Anselmet1984,RB95,Water95,KRS98,Gotoh02,JS2007,SAW2018}. However, a few workers in the recent past \cite{Tang19,Antonia19} have wondered if the deviations from self-similarity should be attributed to insufficiently high Reynolds numbers, and have noted that many measurements and simulations do not satisfy the $4/5$-ths law convincingly enough.

Using a DNS database that spans over a decade in $\rel$ with the highest $\rel=1300$, and by establishing statistical isotropy by decomposing the structure functions in the SO(3) basis, we have shown that the $4/5$-ths law holds to within statistical errors (Fig.~\ref{dlll.fig}). We use SO(3) to remove the lingering effects of anisotropies due to forcing at large scales and the cubic configuration of the computational domain. We have explicitly demonstrated that the second order exponents depart from the Kolmogorov value of $2/3$ and approach a constant value of $0.72 \pm 0.004$ at higher Reynolds numbers. This result soundly demonstrates (the small effect of) intermittency even at the level of the energy spectrum. The fact that the second-order inertial range exponent possesses a constant positive value which initially increases with the Reynolds number has important theoretical implications \cite{Drivas2019}.

We have obtained further results. We have confirmed under convincing conditions that intermittency increases with increasing order for longitudinal, transverse and mixed structure functions, but have shown, quite importantly, that the transverse exponents differ from longitudinal ones for orders greater than about $4$. This result was already obtained \cite{Chen,Dhruva}, but was called into question from symmetry arguments \cite{Procaccia} by stating that longitudinal and transverse structure functions mix different scaling functions and may obscure pure scaling. Be that as it may, the present results show that they are different when they scale. This result suggests the very notion of scaling in turbulence is more complex than traditionally thought.

Lastly, perhaps the most important among the new results, is the finding that the transverse exponents saturate for large moment orders. If there is a mixing of scaling functions, the transverse exponents will control the scaling for very small scales (effectively, very large Reynolds numbers), so the result is fundamentally important. For the highest Reynolds number we have considered here, $\zeta^T_\infty \approx 2$. In the fractal terminology, dimension $2$ indicates the presence of surfaces. It is also the co-dimension of cliffs with unity fractal dimension in the transverse velocity field. The saturation of the exponents suggests that they are controlled by the large jumps that occur in the transverse velocity gradients (Fig.~\ref{traces.fig}). Transverse increments over inertial distances can be obtained by suitably integrating transverse gradients that characterize vorticity. Thus, a consistent and plausible physical picture is the likely prevalence of two-dimensional vortex sheets across which jumps in the transverse velocity field, of the sort seen in Fig.~\ref{traces.fig}(d), might arise. Unfortunately, the direct identification of sheet-like structures in high Reynolds number flows, whose existence is suggested by the saturation of exponents reported here, is a challenging task. This is an ongoing project and will be reported in the future.

The saturation of transverse exponents is the first such confirmation in the general case of homogeneous isotropic turbulence. This raises the possibility that saturation of velocity exponents is an endemic feature of turbulence at high Reynolds number flows. If this is so, the phenomenological small-scale models will have to account for their explicit presence.

\begin{figure}
\begin{minipage}{0.5\textwidth}
\includegraphics [width=1.0\textwidth]{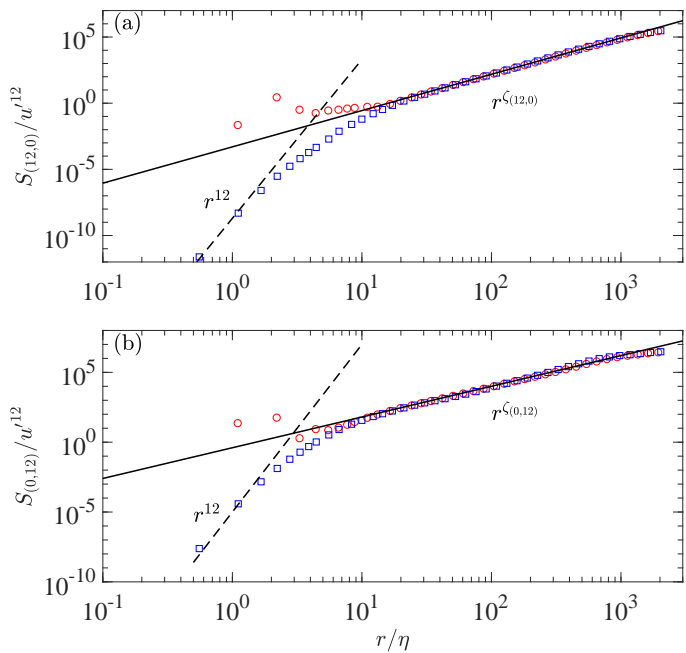}
\end{minipage}
\protect\caption{
Twelfth-order isotropic velocity structure functions versus $r/\eta$ at $\rel=650$ in the (a) longitudinal and (b) transverse directions normalized by the velocity fluctuation $u^\prime$. The two curves in each panel correspond to curves from two DNS at different spatial and temporal resolutions, (circle) $\delxnr = 1.11$, $\deltnr = 0.41$, (square) $\delxnr = 0.55$, $\deltnr = 0.06$. In the inertial range ($100 < r/\eta < 1000$) the structure functions at this order (and below) from both DNS collapse onto each other with the same exponent that is marked by the solid lines (a) $\zeta_{(12,0)} = 2.7 \pm 0.04$ (b) $\zeta_{(0,12)} = \zeta^T_\infty = 2.2 \pm 0.1$; however, only the finer DNS yields the exact $r/\eta \to 0$ exponent shown by the dashed lines.
}
\label{res.fig}
\end{figure}
\section{Acknowledgments}
We are grateful to many of our colleagues with whom we have discussed these results over the years, in one form or another, and to Xiaomeng Zhai and Matthew Clay for their participation in the $16384^3$ computations. This work is partially supported by the National Science Foundation (NSF), via Grants No. ACI-$1640771$  and No. ACI-$1036170$  at the Georgia Institute of Technology. The computations were performed using supercomputing resources provided through the XSEDE consortium (which is funded by NSF) at the Texas Advanced Computing Center at the University of Texas (Austin), and the Blue Waters Project at the National Center for Supercomputing Applications at the University of Illinois (Urbana-Champaign).
\appendix
\section{Numerical resolution}
\label{numres.app}
\noindent
In order to examine the effects of finite spatial and temporal resolution of the DNS we have compared the velocity structure functions from simulations with two different resolutions up to order $12$ at $\rel=650$.  The finer DNS has a spatial resolution $\Delta x$ of almost half the Kolmogorov length scale $\eta$ and a time-step $\Delta t$ which is $6\%$ of the Kolmogorov timescale $\tau_\eta \equiv (\nu/\epsm)^{1/2}$ \cite{YSP18}. In comparison, the coarser DNS has a grid spacing of almost $\eta$ and a time-step which is $40\%$ of $\tau_\eta$. Figure \ref{res.fig} compares the longitudinal and transverse structure functions from two simulations at order $12$ at $\rel=650$. In the range $r/\eta \to 0$, the exponents obtained from a Taylor-series expansion show that only the structure functions from the finer DNS are reliable.
Notwithstanding the small-$r$ result, in the inertial range, the structure functions from both simulations show excellent agreement, as seen in Fig.~\ref{res.fig}, with the same inertial range exponents obtained from least-square fits.

\end{document}